\documentclass[conference]{IEEEtran}
\IEEEoverridecommandlockouts
\usepackage{cite}
\usepackage{amsmath,amssymb,amsfonts}
\usepackage{algorithmic}
\usepackage{graphicx}
\usepackage{textcomp}
\usepackage{multirow}
\usepackage{xcolor}
\def\BibTeX{{\rm B\kern-.05em{\sc i\kern-.025em b}\kern-.08em
    T\kern-.1667em\lower.7ex\hbox{E}\kern-.125emX}}
\begin{document}

\title{HQ-MPSD: A Multilingual Artifact-Controlled Benchmark for Partial Deepfake Speech Detection}

\author{
\IEEEauthorblockN{Menglu Li\IEEEauthorrefmark{1}\IEEEauthorrefmark{2},
Majd Alber\IEEEauthorrefmark{2},
Ramtin Asgarianamiri\IEEEauthorrefmark{3},
Lian Zhao\IEEEauthorrefmark{2},
Xiao-Ping Zhang\IEEEauthorrefmark{1}\IEEEauthorrefmark{2}}
\IEEEauthorblockA{\IEEEauthorrefmark{1}Shenzhen Key Laboratory of Ubiquitous Data Enabling, Tsinghua Shenzhen International Graduate School, Tsinghua University 
}
\IEEEauthorblockA{\IEEEauthorrefmark{2}Department of Electrical, Computer \& Biomedical Engineering, Toronto Metropolitan University, Toronto, Canada\\
Emails: \{menglu.li, majd.alber, ramtin.asgarianamiri, l5zhao\}@torontomu.ca, xpzhang@ieee.org}}
\maketitle

\begin{abstract}
Detecting partial deepfake speech is challenging because manipulations occur only in short regions while the surrounding audio remains authentic. However, existing detection methods are fundamentally limited by the quality of available datasets, many of which rely on outdated synthesis systems and generation procedures that introduce dataset-specific artifacts rather than realistic manipulation cues. To address this gap, we introduce HQ-MPSD, a high-quality multilingual partial deepfake speech dataset. HQ-MPSD is constructed using linguistically coherent splice points derived from fine-grained forced alignment, preserving prosodic and semantic continuity and minimizing audible and visual boundary artifacts. The dataset contains 350.8 hours of speech across eight languages and 550 speakers, with background effects added to better reflect real-world acoustic conditions. MOS evaluations and spectrogram analysis confirm the high perceptual naturalness of the samples.We benchmark state-of-the-art detection models through cross-language and cross-dataset evaluations, and all models experience performance drops exceeding 80\% on HQ-MPSD. These results demonstrate that HQ-MPSD exposes significant generalization challenges once low-level artifacts are removed and multilingual and acoustic diversity are introduced, providing a more realistic and demanding benchmark for partial deepfake detection. The dataset can be found at: \textit{https://zenodo.org/records/17929533}
\end{abstract}


\section{Introduction}
\label{sec:intro}
The rapid progress of speech synthesis has enabled the generation of highly natural artificial speech, which raises growing concerns regarding its misuse in security-critical scenarios \cite{chintha2020recurrent, li24oa_interspeech, alali2025partial}. Among emerging threats, partial deepfake speech poses particular difficulty, where only a portion of an utterance, such as a word or short phrase, is replaced with synthetic speech segments while the surrounding content remains genuine \cite{zhang2021initial}. Because partial deepfakes contain a substantial amount of bonafide speech, they can easily bypass existing detection systems and facilitate misinformation or impersonation \cite{yi2022add, yi2023add, alali2025partial}. Detecting such manipulations is considerably more difficult than detecting fully deepfake speech, as models must localize brief, subtle alterations embedded within an otherwise authentic utterance \cite{li2025survey}. This challenge motivated initiatives such as the ADD 2022 Challenge \cite{yi2022add}, which called for dedicated research on partial deepfake detection.

Despite growing attention, progress in partial deepfake detection is still limited by the scarcity and quality of available datasets. Only a few public resources exist, and many rely on early synthesis systems or simplistic generation strategies that introduce dataset-specific artifacts \cite{negroni2024analyzing}. Models trained on such data may overfit to these superficial cues and generalize poorly to realistic manipulations or unseen acoustic conditions. High-quality datasets are therefore essential to ensure that detectors learn genuine manipulation characteristics rather than artifacts arising from dataset construction.

\begin{figure}

\includegraphics[width=1\linewidth, height=2.3cm]{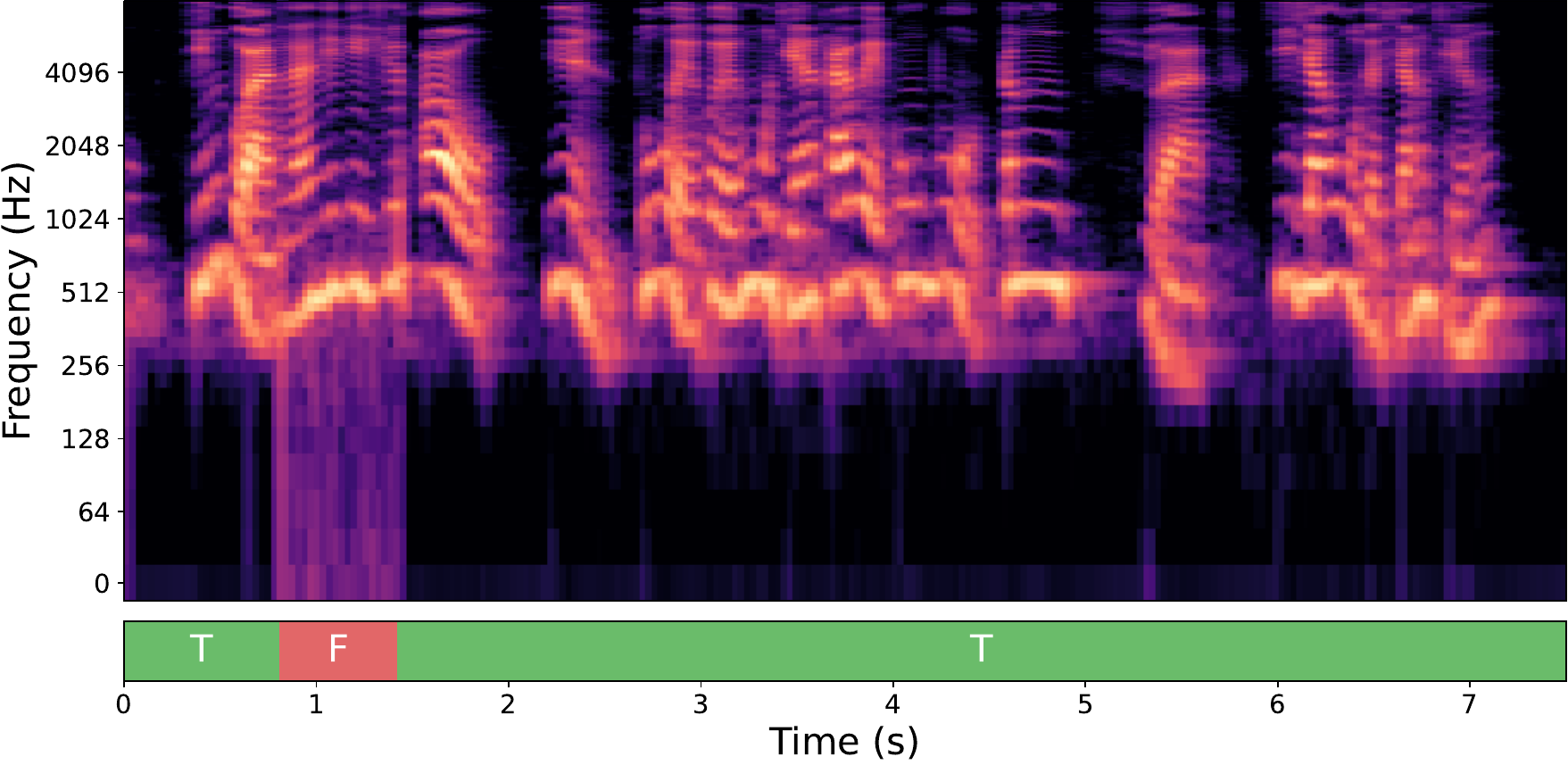} 
\centerline{\begin{minipage} {7cm}
\small (a) HAD\_dev\_fake\_00000006.wav from Half-Truth 
\vspace{0.25cm}
\end{minipage}}

\includegraphics[width=1\linewidth, height=2.3cm]{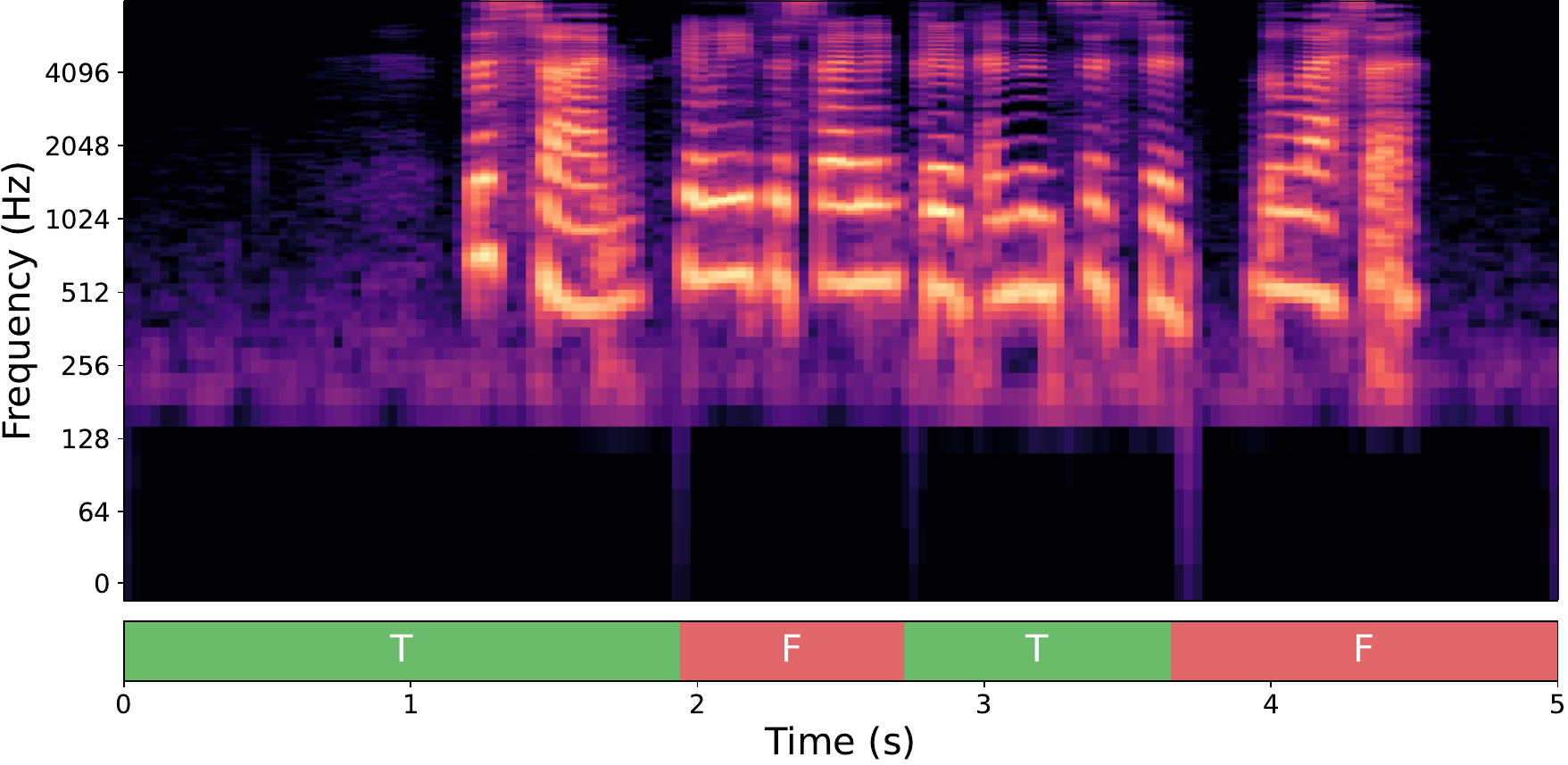}
\centerline{\begin{minipage} {7cm}
\small (b) CON\_D\_0000001.wav from PartialSpoof
\vspace{0.2cm}
\end{minipage}}

\includegraphics[width=1\linewidth, height=2.3cm]{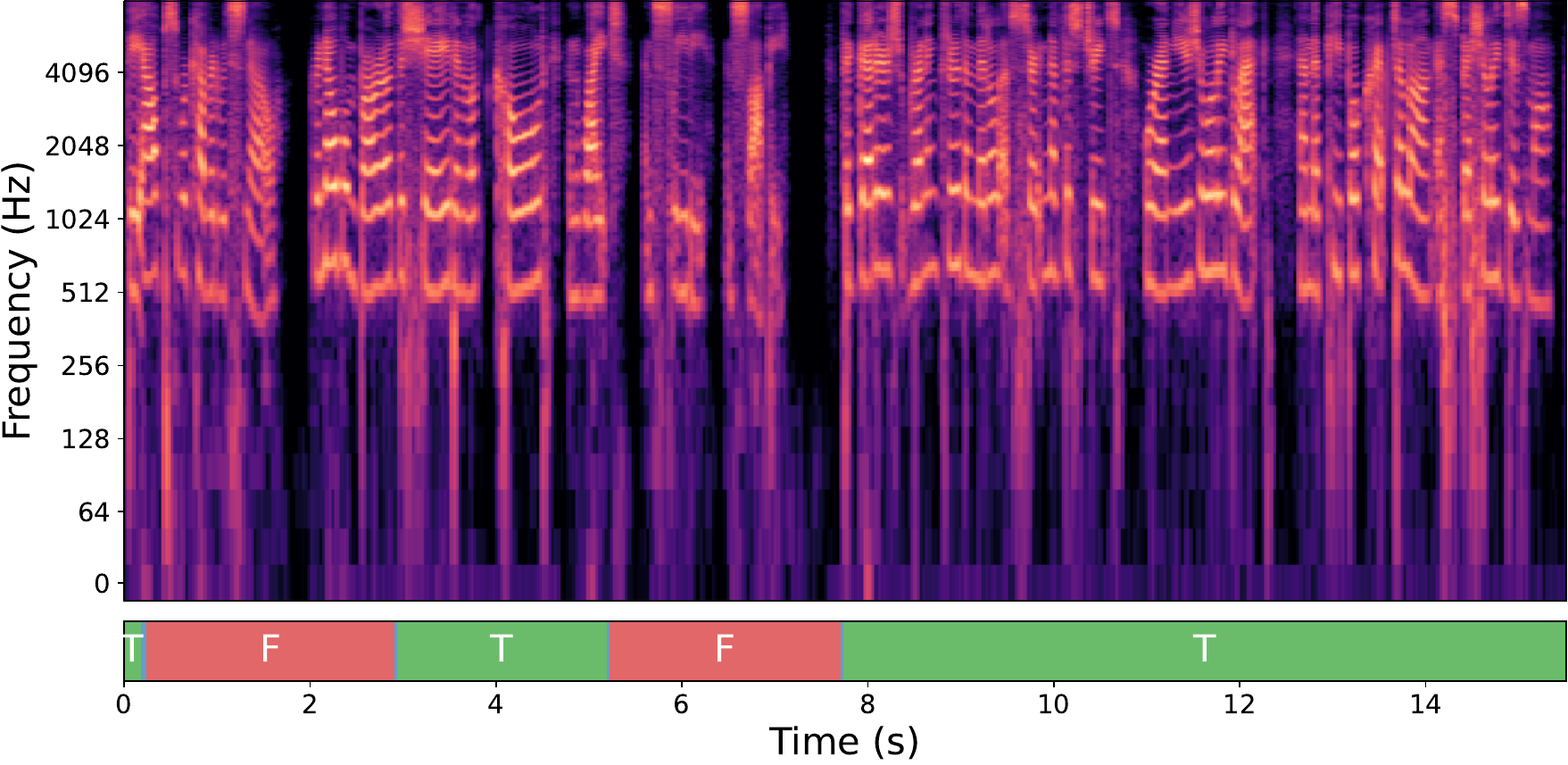}
\centerline{\begin{minipage} {8cm}
\small (c) 2836\_5354\_000070\_000001\_m.flac from our HQ-MPDS
\end{minipage}}

\caption{Mel-spectrograms of partial deepfake speech samples from the Half-Truth, PartialSpoof, and our proposed HQ-MPDS datasets. The colored timeline below each spectrogram indicates the frame-level labels: green denotes bonafide segments, red denotes spoofed segments, and blue (when present) denotes transition regions. While earlier datasets exhibit more distinct visual artifacts at manipulation points, the modifications in HQ-MPDS appear more natural and less visually pronounced.}
\label{fig:spectrogram}
\end{figure}

\begin{table*}[th]
    \caption{ The statistic of our proposed dataset with comparison with existing partial deepfake speech datasets}
    \label{Table:overall}
    \centering
    \setlength\tabcolsep{3pt} 
    \renewcommand{\arraystretch}{1.3}
\begin{tabular}{c|ccccccccc}
\hline
\textbf{}     & \textbf{Year} & \textbf{\# of language} & \textbf{Synthesized Type} & \textbf{Condition} & \textbf{\# of bonafide} & \textbf{\# of deepfake} & \textbf{\# of speakers} & \textbf{Sample Rate} & \textbf{MOS*} \\ \hline
PartialSpoof \cite{zhang2022partialspoof}  & 2021          & 1                       & TTS, VC                   & Clean              & 12483                   & 108978                  & 48                      & 16k Hz  & 3.41 $\pm$ 0.21             \\
Half-Truth \cite{yi21_interspeech}           & 2021          & 1                       & TTS                       & Clean              & 53612                   & 753612                  & 218                     & 44.1k Hz    & 3.43 $\pm$ 0.17         \\
PartialEdit \cite{zhang25g_interspeech}  & 2025          & 1                       & Natural Codec             & Clean              & -                       & 43358                   & 108                     & 16k Hz     &    3.41 $\pm$ 0.21       \\ \hline
\textbf{HQ-MPDS (Ours)} & 2025          & \textbf{8}                       & TTS, VC                       & \textbf{Noise, RIR}         & 51715                   & 103430                   & \textbf{550}                     & 16k Hz            & \textbf{3.68} $\pm$ 0.12  \\\hline
\end{tabular}
          \begin{minipage}{0.9\textwidth} 
          
          \vspace{2pt}
        \linespread{0.6}\selectfont
        {\scriptsize  $^{*}$MOS evaluation is performed exclusively on the available partial deepfake speech samples within each dataset.\\}
        \end{minipage}
\end{table*}

Existing partial deepfake datasets exhibit three key limitations.
(1) \textbf{Low sample quality}. Most datasets create partial deepfakes by concatenating randomly selected bonafide and deepfake segments without ensuring speaker consistency or acoustic compatibility. This often produces unnatural transitions, inconsistent speaker characteristics, and clear splicing artifacts that are easily visible in mel-spectrograms, as shown in Fig. \ref{fig:spectrogram}(a) and (b) Simple frequency-based detectors can exploit these artifacts to achieve satisfactory accuracy, indicating that they may learn dataset-specific flaws rather than actual manipulation cues.
(2) \textbf{Insufficient utterance length}. Many partial deepfake speech samples, particularly in PartialSpoof \cite{zhang2022partialspoof}, are shorter than 5 seconds, with some under 1 second. Such brief clips lack meaningful phonetic or prosodic structure and limit a model’s ability to capture contextual or long-range dependencies critical for detecting subtle manipulations.
(3) \textbf{Limited generalization capability}. Existing datasets are predominantly monolingual and created under clean laboratory conditions, whereas real-world speech varies substantially across languages, accents, and acoustic environments. Models trained under these constrained settings tend to overfit language- or noise-specific patterns, which can lead to severe degradation when evaluated in cross-lingual or noisy scenarios.

To address these limitations, we introduce HQ-MPSD, a high-quality multilingual partial deepfake speech dataset designed to support robust and generalizable deepfake detection research. HQ-MPSD contains 350.8 hours of both fully and partially manipulated speech across eight languages. Each bonafide–deepfake pair is acoustically aligned through loudness and spectral normalization, and partial manipulations are created using linguistically coherent splice points derived from word-level forced alignment. These design choices preserve prosodic and semantic continuity while minimizing boundary artifacts that could otherwise be exploited by detectors. Furthermore, background effects are applied to partial deepfake samples to reduce clean-lab bias and mask superficial background mismatches between bonafide and synthesized segments. A key novelty of HQ-MPDS is that both audible and visual splicing artifacts are substantially reduced, so that producing manipulated segments that cannot be trivially exposed through mel-spectrogram inspection or simple heuristics. Utterance lengths are constrained to 5–15 seconds to provide linguistically meaningful contexts, and Mean Opinion Score (MOS) evaluations confirm the high perceptual naturalness of speech samples.

To assess the challenges posed by HQ-MPSD, we conduct two sets of experiments. First, we examine cross-language generalization, evaluating whether state-of-the-art models trained on English extend effectively to seven additional languages. Second, we evaluate cross-dataset generalization, testing whether models trained on existing partial deepfake datasets transfer to the high-quality, artifact-controlled conditions presented by HQ-MPSD. Across both settings, model performance degrades sharply, revealing substantial generalization gaps once superficial artifacts are removed and multilingual and acoustic variability are introduced. These findings position HQ-MPSD as a multilingual, artifact-controlled benchmark that addresses limitations of prior datasets and aims to facilitate the development of detection models that learn genuine manipulation cues for reliable open-world performance.

\section{Related Work}
\label{sec:related}
\subsection{Partial Deepfake Detection Techniques}
Partial deepfake speech detection methods generally fall into three categories: frame-level classification, multi-task learning, and boundary detection. Frame-level methods \cite{liu2023transsionadd, liu2024harmonet} divide an utterance into short segments and classify each independently. While simple and straightforward, their performance depends heavily on precise temporal labels and they often struggle with short or ambiguous frames. Multi-task learning approaches \cite{li2023convolutional, li2023multi} combine frame-level and utterance-level objectives to improve robustness, but the need to jointly optimize multiple predictors increases architectural complexity and makes the training process sensitive to label noise. Boundary detection models \cite{cai2024integrating, liu24m_interspeech} aim to identify the transition between bonafide and manipulated regions. These models perform well when transitions exhibit clear acoustic cues but may focus on dataset-specific discontinuities rather than true synthesis artifacts, which may limit their generalization to more natural or subtle manipulations.

Overall, existing methods are fundamentally constrained by the characteristics of the datasets they are trained on. Accurate modeling of partial manipulations requires datasets with consistent acoustic conditions, high perceptual quality, and fine-grained temporal annotations. Without these properties, models can overfit to superficial dataset-related artifacts and fail to generalize to realistic scenarios.

\subsection{Existing Datasets}
There is a limited number of publicly available datasets dedicated to partial deepfake speech. PartialSpoof \cite{zhang2022partialspoof} is the first to introduce the concept by generating samples through random swapping of short segments between bonafide and fully deepfake utterances. Although simple, this strategy often breaks linguistic continuity and produces clear signal discontinuities that are easy to detect through spectral analysis. Models trained on such data risk overfitting to these splicing artifacts rather than learning true manipulation cues. Half-Truth \cite{yi21_interspeech}, the first Chinese dataset, applies a similar swapping strategy and likewise ignores speaker consistency and transition smoothness. This results in acoustically mismatched and semantically incoherent utterances that limits its ability to represent natural speech transitions. There are some dataset introduced recently. PartialEdit \cite{zhang25g_interspeech} focuses on neural codec–based editing, while SynSpeech \cite{qiu2025synspeech} and LlamaPartialSpoof \cite{luong2025llamapartialspoof} incorporate modern speech synthesis techniques. However, these datasets remain monolingual, primarily clean, and do not address diverse acoustic environments or controlled artifact settings for partial deepfake generation.

These datasets collectively highlight the need for more realistic resources that support natural transition, consistent speaker identity, and broader linguistic and acoustic diversity. HQ-MPSD aims to address this gap by aligning content between bonafide and deepfake speech, smoothing transitions through linguistically coherent replacements, and incorporating multilingual and acoustically varied conditions. This design provides a more reliable benchmark for evaluating model generalization and encourages the development of detection systems that focus on intrinsic manipulation cues rather than artifacts introduced during data construction.

\section{The Proposed Dataset}
\label{sec:dataset}
\begin{figure}
\centering{\includegraphics[width=0.98\columnwidth]{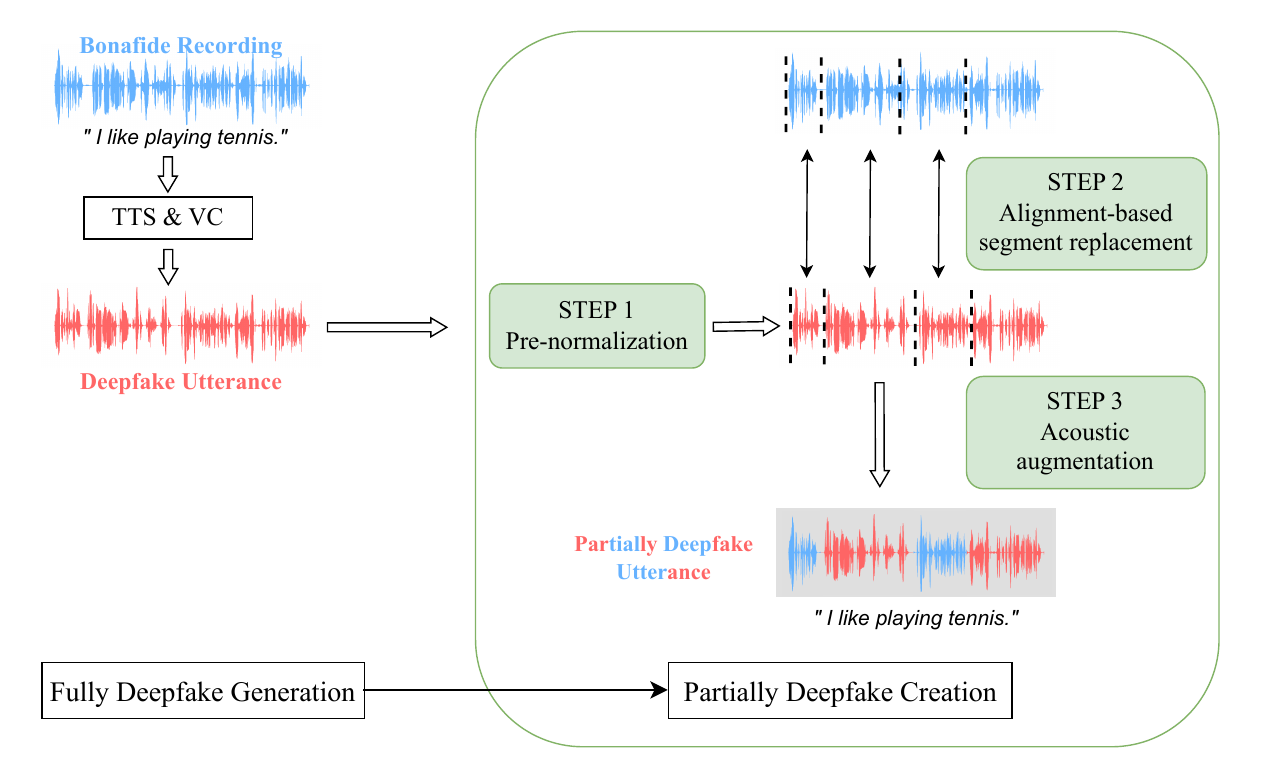}}
\caption{The generation pipeline of our proposed dataset. Fully Deepfake Generation uses TTS and VC models to synthesize complete utterances. Partially Deepfake Creation consists of three steps: (1) Normalization, including loudness and spectral brightness adjustment; (2) Word-level Forced Alignment to determine precise splicing boundaries; and (3) Background Effect Augmentation using room impulse responses and/or noise to blend the partial deepfake speech with realistic environmental effects.}
\label{fig:pipeline}
\end{figure}

This section introduces HQ-MPSD and outlines the key components of its generation pipeline, along with the properties that make it a comprehensive benchmark for evaluating deepfake detection under realistic and diverse conditions. The overall pipeline is illustrated in Fig. \ref{fig:pipeline}.

\subsection{Fully Deepfake Speech Generation}
The fully deepfake subset is built from the Multilingual LibriSpeech corpus \cite{Pratap_Xu_Sriram_Synnaeve_Collobert_2020}, which provides transcribed long-form audiobook recordings in eight languages: Dutch, English, French, German, Italian, Polish, Portuguese, and Spanish. Long-form recordings are segmented into 5–15 s utterances and paired with their transcripts. XTTSv2 \cite{Eren_Coqui_TTS_2021} is used to synthesize deepfake speech by conditioning on each utterance’s transcript and its corresponding bonafide audio as the reference voice. This produces speaker-matched and linguistically aligned synthetic speech across all languages. Multiple speakers are selected per language to ensure diversity in accent, style, and timbre. The one-to-one mapping between bonafide and deepfake utterances provides a clean foundation for controlled partial manipulation.

\subsection{Partial Deepfake Speech Creation}
We generate high-quality partially manipulated utterances through a three-stage process designed to preserve acoustic coherence and natural prosody.

\textbf{Step 1: Pre-normalization} 
Before replacement, we normalize the loudness and spectral balance between bonafide and deepfake speech to reduce superficial mismatches. RMS-based loudness alignment together with adaptive pre-emphasis filtering mitigates loudness and spectral disparities, particularly the spectral imbalance commonly introduced by neural vocoders, while preserving speaker identity. This step ensures that segment replacement is not driven by trivial acoustic differences but instead reflects meaningful synthesis artifacts.

\textbf{Step 2: Alignment-based segment replacement} 
Following preprocessing, we generate partial deepfake speech by replacing selected segments in bonafide utterances with the corresponding portions from their normalized deepfake counterparts. Unlike simple timestamp-based or Voice Activity Detection-based cutting, which often disrupts prosody and introduces unnatural discontinuities, our approach determines linguistically coherent swap points using word-level forced alignment. Each bonafide-deepfake pair is first transcribed using Whisper \cite{Bain_Huh_Han_Zisserman_2023}, and only pairs with closely matching transcripts are retained, which also serves as an additional verification of synthesis quality. Forced alignment is then obtained using the Montreal Forced Aligner \cite{McAuliffe_Socolof_Mihuc_Wagner_Sonderegger_2017}, and replacement boundaries are placed at midpoints between aligned word pairs to avoid cutting across phones or prosodic transitions. A limited number of segments are substituted per utterance, and all boundaries are smoothed with a 30 ms overlap-add using cosine fading to remove clicks and ensure seamless acoustic transitions. This alignment strategy produces mixtures that preserve natural prosody and achieve high perceptual consistency, which outperforms approaches that rely on coarse or unconstrained cuts.

\textbf{Step 3: Acoustic augmentation} 
To introduce environmental diversity and better reflect real-world recording conditions, we apply noise and reverberation to the generated partial deepfake utterances. Room acoustics are simulated by convolving each waveform with a randomly selected room impulse response from OpenSLR 26 \cite{Ko_Peddinti_Povey_Seltzer_Khudanpur_2017}, and background noise from MUSAN \cite{snyder2015musan} is added at 15 dB SNR. Different combinations of reverberation and noise are used to create varied acoustic scenarios. This augmentation step extends the dataset beyond clean studio-style recordings and produces samples that more closely resemble practical usage conditions.

\subsection{Multi-level Labeling}
Each sample is assigned an utterance-level label indicating whether it is bonafide, fully deepfake, or partial deepfake. For partially manipulated utterances, we further provide frame-level annotations using non-overlapping 30 ms frames. Each frame is labeled as bonafide, deepfake, or transition, where transition frames correspond to regions near swap boundaries affected by crossfading. Including a dedicated transition label helps separate boundary artifacts from genuine synthesis cues, which may enable a clearer interpretation of model behavior and support more accurate evaluation of fine-grained detection performance.

\begin{figure}
\centering{\includegraphics[width=\columnwidth]{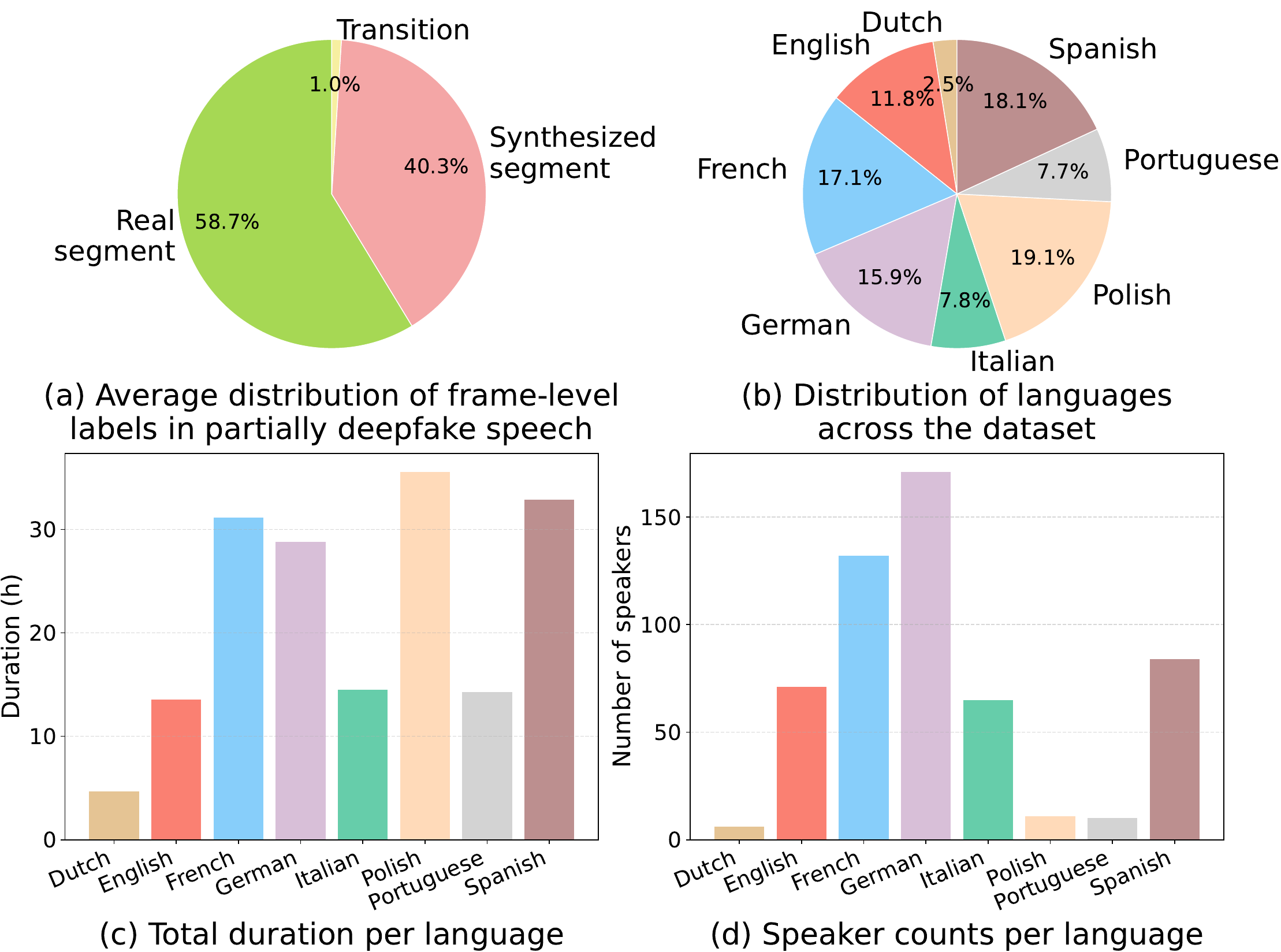}}
\caption{Overview statistics of our proposed HQ-MPDS dataset.}
\label{fig:overall}
\end{figure}

\subsection{Dataset Statistic Information}
HQ-MPSD contains eight language subsets, each following a unified processing pipeline. The overall language distribution, total duration, and number of speakers per language are shown in Fig. \ref{fig:overall}. In total, the dataset includes 550 speakers and approximately 155k utterances, each ranging from 5 to 15 seconds in duration. For every linguistic instance, we provide a matched triplet of bonafide, fully deepfake, and partial deepfake samples, with additional variants that include neutral background effects. This structure offers consistent linguistic alignment across conditions and supports controlled comparisons in downstream evaluation.

Table \ref{Table:overall} summarizes the statistics of our proposed dataset. To the best of our knowledge, HQ-MPDS is the first dataset that offers multilingual coverage and explicitly incorporates neutral background effects. To assess perceptual quality, we apply DNSMOS \cite{reddy2021dnsmos} to the partial deepfake speech samples. HQ-MPSD achieves an average MOS of 3.58, which represents the highest naturalness level among the existing datasets.

\subsection{Dataset Properties}
HQ-MPSD possesses several characteristics that make it a valuable benchmark for open-world partial deepfake detection.

\textbf{Multilingual diversity.} 
HQ-MPSD includes speech samples in eight languages, each with multiple speakers with varied genders and age groups. This multilingual composition introduces substantial phonetic and prosodic variability. It allows comprehensive cross-lingual evaluation and provides a strong benchmark for assessing model generalization across diverse linguistic contexts.

\textbf{High perceptual quality.}
HQ-MPSD achieves high MOS scores and maintains clear transcript fidelity. Whisper-based transcription consistency further verifies the clarity of synthesized speech. Fig. \ref{fig:spectrogram}(c) illustrates the mel-spectrogram of a partial deepfake sample from HQ-MPDS with corresponding frame-level annotations. The partially manipulated spectrograms exhibit smooth transitions without visible discontinuities. Furthermore,the inclusion of background effects helps to mask potential discontinuities across bonafide and manipulated regions. This design minimizes concatenation artifacts and enhances the overall perceptual quality of the speech samples.

\textbf{Fine-grained paired structure.}
Each linguistic instance is provided as a paired set containing a bonafide recording, a fully deepfake version, and a partial deepfake version, with background-effect variants. All versions share identical linguistic content and alignment. This fine-grained structure supports controlled comparisons at both the utterance and segment levels and enables detailed analysis of how manipulation cues influence detection models.

\section{Experiment}
Although the primary contribution of this work lies in the construction of HQ-MPSD, it is equally important to demonstrate the open-world challenges revealed by this dataset. We therefore conduct two sets of experiments: (1) cross-language evaluation to assess multilingual generalization, and (2) cross-dataset evaluation to test whether models trained on existing partial deepfake datasets generalize to the conditions presented in HQ-MPSD. Overall, the experiments demonstrate that HQ-MPSD reveals critical generalization gaps in existing detection models.

\begin{table*}[th]
    \caption{Cross-language evaluation of the baseline systems in intra-lingual settings (English) and cross-lingual settings across seven unseen languages. The best and second-best results in each category are shown in \textbf{bold} and \underline{underline}, respectively.}
    \label{Table:experiment}
    \centering
    \setlength\tabcolsep{3pt}
    \renewcommand{\arraystretch}{1.3}
    \begin{tabular}{cc|cc|cccccccccccccc}
\hline
\multicolumn{2}{c|}{\multirow{2}{*}{\textbf{Baseline Models}}} & \multicolumn{2}{c|}{\textbf{Intra-Lingual}}         & \multicolumn{14}{c}{\textbf{Cross-Lingual Performance}}                                                                                                                                                                                                                                                                                                                                                                                                                                                                      \\ \cline{3-18} 
\multicolumn{2}{c|}{}                                          & \multicolumn{2}{c|}{English}                        & \multicolumn{2}{c|}{\textbf{French}}                                      & \multicolumn{2}{c|}{\textbf{Polish}}                                      & \multicolumn{2}{c|}{\textbf{German}}                                      & \multicolumn{2}{c|}{\textbf{Spanish}}                                     & \multicolumn{2}{c|}{\textbf{Italian}}                                     & \multicolumn{2}{c|}{\textbf{Portuguese}}                                  & \multicolumn{2}{c}{\textbf{Dutch}}                   \\ \hline
\multicolumn{1}{c|}{Feature}         & Classifier     & \multicolumn{1}{c|}{EER$\downarrow$}             & AUC$\uparrow$             & \multicolumn{1}{c|}{EER$\downarrow$}            & \multicolumn{1}{c|}{AUC$\uparrow$ }            & \multicolumn{1}{c|}{EER$\downarrow$}            & \multicolumn{1}{c|}{AUC$\uparrow$}            & \multicolumn{1}{c|}{EER$\downarrow$}            & \multicolumn{1}{c|}{AUC$\uparrow$ }            & \multicolumn{1}{c|}{EER$\downarrow$}            & \multicolumn{1}{c|}{AUC$\uparrow$ }            & \multicolumn{1}{c|}{EER$\downarrow$}            & \multicolumn{1}{c|}{AUC$\uparrow$ }            & \multicolumn{1}{c|}{EER$\downarrow$}            & \multicolumn{1}{c|}{AUC$\uparrow$ }            & \multicolumn{1}{c|}{EER$\downarrow$}            & AUC$\uparrow$             \\ \hline
\multicolumn{1}{c|}{SincNet}              & GAT-ST             & \multicolumn{1}{c|}{0.88}          & 0.988          & \multicolumn{1}{c|}{40.25}          & \multicolumn{1}{c|}{0.554}          & \multicolumn{1}{c|}{42.55}          & \multicolumn{1}{c|}{0.451}          & \multicolumn{1}{c|}{39.29}          & \multicolumn{1}{c|}{0.517}          & \multicolumn{1}{c|}{40.98}          & \multicolumn{1}{c|}{0.518}          & \multicolumn{1}{c|}{36.79}          & \multicolumn{1}{c|}{0.503}          & \multicolumn{1}{c|}{36.81}          & \multicolumn{1}{c|}{0.632}          & \multicolumn{1}{c|}{46.83}          & 0.444          \\
\multicolumn{1}{c|}{MFCC}                 & GAT-ST             & \multicolumn{1}{c|}{4.28}          & 0.976          & \multicolumn{1}{c|}{\textbf{28.83}} & \multicolumn{1}{c|}{\textbf{0.782}} & \multicolumn{1}{c|}{\textbf{21.64}} & \multicolumn{1}{c|}{\textbf{0.857}} & \multicolumn{1}{c|}{\textbf{27.82}} & \multicolumn{1}{c|}{\textbf{0.788}} & \multicolumn{1}{c|}{\underline{31.41}}    & \multicolumn{1}{c|}{\underline{0.752}}    & \multicolumn{1}{c|}{\underline{28.04}}    & \multicolumn{1}{c|}{\underline{0.785}}    & \multicolumn{1}{c|}{\textbf{23.37}} & \multicolumn{1}{c|}{\textbf{0.834}} & \multicolumn{1}{c|}{\textbf{28.67}} & \textbf{0.786} \\
\multicolumn{1}{c|}{Spectrogram}          & GAT-ST             & \multicolumn{1}{c|}{2.51}          & 0.978          & \multicolumn{1}{c|}{43.24}          & \multicolumn{1}{c|}{0.591}          & \multicolumn{1}{c|}{49.42}          & \multicolumn{1}{c|}{0.494}          & \multicolumn{1}{c|}{47.16}          & \multicolumn{1}{c|}{0.538}          & \multicolumn{1}{c|}{44.36}          & \multicolumn{1}{c|}{0.576}          & \multicolumn{1}{c|}{42.62}          & \multicolumn{1}{c|}{0.598}          & \multicolumn{1}{c|}{37.12}          & \multicolumn{1}{c|}{0.671}          & \multicolumn{1}{c|}{55.31}          & 0.410          \\
\multicolumn{1}{c|}{W2v2-XLSR}             & GAT-ST             & \multicolumn{1}{c|}{\underline{0.59}}    & \underline{0.995}    & \multicolumn{1}{c|}{42.91}          & \multicolumn{1}{c|}{0.613}          & \multicolumn{1}{c|}{43.89}          & \multicolumn{1}{c|}{0.498}          & \multicolumn{1}{c|}{42.03}          & \multicolumn{1}{c|}{0.627}          & \multicolumn{1}{c|}{43.19}          & \multicolumn{1}{c|}{0.602}          & \multicolumn{1}{c|}{44.32}          & \multicolumn{1}{c|}{0.592}          & \multicolumn{1}{c|}{41.03}          & \multicolumn{1}{c|}{0.653}          & \multicolumn{1}{c|}{41.87}          & 0.644          \\
\multicolumn{2}{c|}{TDAM}                                      & \multicolumn{1}{c|}{\textbf{0.29}} & \textbf{0.998} & \multicolumn{1}{c|}{\underline{29.47}}    & \multicolumn{1}{c|}{\underline{0.751}}    & \multicolumn{1}{c|}{\underline{36.13}}    & \multicolumn{1}{c|}{\underline{0.675}}    & \multicolumn{1}{c|}{\underline{32.65}}    & \multicolumn{1}{c|}{\underline{0.690}}    & \multicolumn{1}{c|}{\textbf{28.88}} & \multicolumn{1}{c|}{\textbf{0.748}} & \multicolumn{1}{c|}{\textbf{27.49}} & \multicolumn{1}{c|}{\textbf{0.767}} & \multicolumn{1}{c|}{\underline{30.72}}    & \multicolumn{1}{c|}{\underline{0.768}}    & \multicolumn{1}{c|}{\underline{32.57}}    & \underline{0.720}    \\ \hline
\end{tabular}

\end{table*}

\subsection{Cross-Language Performance}
\subsubsection{Baseline Systems}
We evaluate a representative set of widely-used and state-of-the-art (SOTA) systems under multilingual conditions. GAT-ST \cite{Tak_Jung_Patino_Kamble_Todisco_Evans_2021} is adopted as a strong baseline, using graph attention networks as the classifier along with SincNet \cite{ravanelli2018speaker}, MFCC, spectrogram, and W2v2-XLSR \cite{baevski2020wav2vec} front-end features. We also include TDAM \cite{li2025frame}, a recent end-to-end model specifically designed to capture segment-level inconsistencies in partial deepfake speech.

\subsubsection{Experiment Setup }
The English subset of HQ-MPSD is divided into training, validation, and evaluation sets using an 8:1:1 split with no speaker overlap. All models are trained on the English training set and selected based on the best validation loss. Evaluation is performed on the English test set for intra-lingual performance and on seven additional languages for cross-lingual generalization. The task is framed as binary classification between bonafide utterances and utterances containing injected deepfake segments. Models are trained using Adam optimizer with a batch size of 10, and utterances within each batch are zero-padded to avoid truncating manipulated regions.  Learning rates are set to $10^{-3}$ for MFCC, SincNet, and spectrogram features, and $10^{-5}$ for the remaining configurations. 

Performance is reported using Equal Error Rate (EER) and Area Under the Curve (AUC), where lower EER and higher AUC indicate better detection capability.

\subsubsection{Result and Discussion}
Table \ref{Table:experiment} presents benchmark results in both intra-lingual and cross-lingual evaluation settings.

\textbf{Intra-Lingual Performance}
TDAM achieves the strongest performance with an EER of 0.29\% and an AUC of 0.998, which aligns with its design for partial deepfake detection. Among GAT-based systems, models with learnable front-ends, such as W2v2-XLSR and SincNet, outperform the handcrafted spectral features. This indicates that adaptive feature learning effectively captures the fine-grained temporal–spectral inconsistencies present in partial manipulations when the training and testing languages are aligned.

\textbf{Cross-Lingual Performance}
Performance drops considerably for all systems when evaluated on unseen languages. Spectrogram and SincNet features show the most severe degradation, as both are heavily influenced by language-specific phonetic and acoustic characteristics that do not transfer across languages. Notably, W2v2-XLSR, although pretrained on 128 languages, also exhibits poor cross-language robustness once fine-tuned exclusively on English. This suggests overspecialization to the fine-tuning domain. In contrast, MFCC-GAT and TDAM achieve comparatively stronger generalization. MFCC compresses the spectral envelope and removes fine phonetic details, while TDAM emphasizes temporal irregularities that are less dependent on language structure. 

These findings reveal that even SOTA systems struggle to generalize across languages in partial deepfake scenarios, which highlights the urgent need for multilingual datasets such as HQ-MPSD to drive progress toward language-agnostic detection models.

\subsection{Cross-Dataset Performance}

\begin{figure}
\centering{\includegraphics[width=0.95\columnwidth]{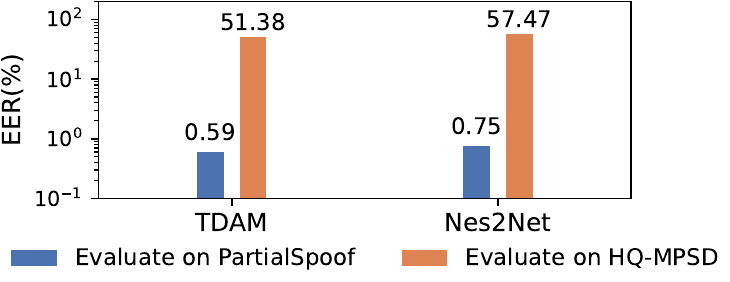}}
\caption{Cross-dataset evaluation of two detection models trained on PartialSpoof, tested on both the PartialSpoof evaluation set and our HD-MPSD English subset. Performance on HD-MPSD shows an increase in EER of up to 90\% compared with the PartialSpoof evaluation.}
\label{fig:comparison}
\end{figure}  
A major motivation brought by HQ-MPSD is the improvement of speech sample quality through the removal of concentrated injection artifacts. To evaluate whether the models’ learned representations can transfer across datasets, we conduct a cross-dataset experiment in a monolingual environment, where models trained on PartialSpoof are tested on the HQ-MPSD English set. TDAM \cite{li2025frame} and Nes2Net \cite{Liu_Truong_Kumar} are selected due to their strong performance on PartialSpoof. Both models utilize W2v2-XLSR to extract deep embedding and are trained with Adam at an initial learning rate of $5 \times 10^{-5}$. Variable-length inputs are handled through batch-wise zero-padding following \cite{li2025frame}, and EER is used as the evaluation metric.

\textbf{Result and Discussion}
Figure \ref{fig:comparison} compares their performance on the in-domain PartialSpoof evaluation set and the out-of-domain HQ-MPSD English subset. Both models show a drastic performance collapse when transferred to HQ-MPSD, with TDAM and Nes2Net reaching EERs of 51.38\% and 57.47\%, respectively, which are worse than random guessing. This sharp degradation demonstrates that existing systems rely heavily on dataset-specific artifacts, such as unnatural boundary cues, which are no longer present in HQ-MPSD. Once these superficial cues are substantially reduced, the models fail to detect genuine manipulation traces.

\section{Conclusion}
We introduce HQ-MPSD, a high-quality multilingual partial deepfake speech dataset comprising 155,145 utterances across eight languages. The dataset is constructed through a carefully designed generation pipeline: a pre-normalization stage aligns loudness and spectral characteristics between bonafide and synthetic speech, and fine-grained forced alignment is then used to select linguistically coherent splice points that preserve prosodic and semantic continuity. These steps, together with the incorporation of neutral background effects, substantially reduce audible and visual boundary artifacts and produce samples that better reflect real-world acoustic conditions. Mel-spectrogram analysis and MOS evaluations further confirm the high perceptual naturalness of the dataset. By suppressing superficial cues and ensuring acoustic consistency, HQ-MPSD encourages detection models to focus on genuine synthesis artifacts rather than dataset-induced patterns.

Using HQ-MPSD, we conduct cross-language and cross-dataset evaluations on SOTA models. When trained on English and tested on other languages, or when transferred from existing datasets to HQ-MPSD under monolingual settings, model performance drops sharply, in some cases degrading toward random guessing. These results reveal that once low-level artifacts are removed and multilingual and acoustic variability are introduced, current detection systems exhibit significant generalization weaknesses. HQ-MPSD therefore serves as a rigorous benchmark and a foundation for developing more robust and generalizable detection methods.


\begin{thebibliography}{10}

\bibitem{chintha2020recurrent}
Akash Chintha, Bao Thai, Saniat~Javid Sohrawardi, Kartavya Bhatt, Andrea Hickerson, Matthew Wright, and Raymond Ptucha,
\newblock ``Recurrent convolutional structures for audio spoof and video deepfake detection,''
\newblock {\em IEEE Journal of Selected Topics in Signal Processing}, vol. 14, no. 5, pp. 1024--1037, 2020.

\bibitem{li24oa_interspeech}
Menglu Li and Xiao-Ping Zhang,
\newblock ``{Interpretable Temporal Class Activation Representation for Audio Spoofing Detection},''
\newblock in {\em {Interspeech 2024}}, 2024, pp. 1120--1124.

\bibitem{alali2025partial}
Abdulazeez Alali and George Theodorakopoulos,
\newblock ``Partial fake speech attacks in the real world using deepfake audio,''
\newblock {\em Journal of Cybersecurity and Privacy}, vol. 5, no. 1, pp. 6, 2025.

\bibitem{zhang2021initial}
Lin Zhang, Xin Wang, Erica Cooper, Junichi Yamagishi, Jose Patino, and Nicholas Evans,
\newblock ``An initial investigation for detecting partially spoofed audio,''
\newblock {\em arXiv preprint arXiv:2104.02518}, 2021.

\bibitem{yi2022add}
Jiangyan Yi, Ruibo Fu, Jianhua Tao, Shuai Nie, Haoxin Ma, Chenglong Wang, Tao Wang, Zhengkun Tian, Ye~Bai, Cunhang Fan, et~al.,
\newblock ``Add 2022: the first audio deep synthesis detection challenge,''
\newblock in {\em ICASSP 2022-2022 IEEE International Conference on Acoustics, Speech and Signal Processing (ICASSP)}. IEEE, 2022, pp. 9216--9220.

\bibitem{yi2023add}
Jiangyan Yi, Jianhua Tao, Ruibo Fu, Xinrui Yan, Chenglong Wang, Tao Wang, Chu~Yuan Zhang, Xiaohui Zhang, Yan Zhao, Yong Ren, et~al.,
\newblock ``Add 2023: the second audio deepfake detection challenge,''
\newblock {\em arXiv preprint arXiv:2305.13774}, 2023.

\bibitem{li2025survey}
Menglu Li, Yasaman Ahmadiadli, and Xiao-Ping Zhang,
\newblock ``A survey on speech deepfake detection,''
\newblock {\em ACM Computing Surveys}, 2025.

\bibitem{negroni2024analyzing}
Viola Negroni, Davide Salvi, Paolo Bestagini, and Stefano Tubaro,
\newblock ``Analyzing the impact of splicing artifacts in partially fake speech signals,''
\newblock {\em arXiv preprint arXiv:2408.13784}, 2024.

\bibitem{zhang2022partialspoof}
Lin Zhang, Xin Wang, Erica Cooper, Nicholas Evans, and Junichi Yamagishi,
\newblock ``The partialspoof database and countermeasures for the detection of short fake speech segments embedded in an utterance,''
\newblock {\em IEEE/ACM Transactions on Audio, Speech, and Language Processing}, vol. 31, pp. 813--825, 2022.

\bibitem{yi21_interspeech}
Jiangyan Yi, Ye~Bai, Jianhua Tao, Haoxin Ma, Zhengkun Tian, Chenglong Wang, Tao Wang, and Ruibo Fu,
\newblock ``Half-truth: A partially fake audio detection dataset,''
\newblock in {\em Interspeech 2021}, 2021, pp. 1654--1658.

\bibitem{zhang25g_interspeech}
You Zhang, Baotong Tian, Lin Zhang, and Zhiyao Duan,
\newblock ``{PartialEdit: Identifying Partial Deepfakes in the Era of Neural Speech Editing },''
\newblock in {\em {Interspeech 2025}}, 2025, pp. 5353--5357.

\bibitem{liu2023transsionadd}
Jie Liu, Zhiba Su, Hui Huang, Caiyan Wan, Quanxiu Wang, Jiangli Hong, Benlai Tang, and Fengjie Zhu,
\newblock ``Transsionadd: A multi-frame reinforcement based sequence tagging model for audio deepfake detection,''
\newblock {\em Proceedings of IJCAI 2023 Workshop on Deepfake Audio Detection and Analysis}, pp. 113--118, 2023.

\bibitem{liu2024harmonet}
Liwei Liu, Huihui Wei, Dongya Liu, and Zhonghua Fu,
\newblock ``Harmonet: Partial deepfake detection network based on multi-scale harmof0 feature fusion,''
\newblock in {\em Proc. Interspeech}, 2024, vol. 2024, pp. 2255--2259.

\bibitem{li2023convolutional}
Kang Li, Xiao-Min Zeng, Jian-Tao Zhang, and Yan Song,
\newblock ``Convolutional recurrent neural network and multitask learning for manipulation region location.,''
\newblock in {\em Proceedings of IJCAI 2023 Workshop on Deepfake Audio Detection and Analysis}, 2023, pp. 18--22.

\bibitem{li2023multi}
Jun Li, Lin Li, Mengjie Luo, Xiaoqin Wang, Shushan Qiao, and Yumei Zhou,
\newblock ``Multi-grained backend fusion for manipulation region location of partially fake audio.,''
\newblock in {\em Proceedings of IJCAI 2023 Workshop on Deepfake Audio Detection and Analysis}, 2023, pp. 43--48.

\bibitem{cai2024integrating}
Zexin Cai and Ming Li,
\newblock ``Integrating frame-level boundary detection and deepfake detection for locating manipulated regions in partially spoofed audio forgery attacks,''
\newblock {\em Computer Speech \& Language}, vol. 85, pp. 101597, 2024.

\bibitem{liu24m_interspeech}
Tianchi Liu, Lin Zhang, Rohan~Kumar Das, Yi~Ma, Ruijie Tao, and Haizhou Li,
\newblock ``How do neural spoofing countermeasures detect partially spoofed audio?,''
\newblock in {\em Interspeech 2024}, 2024, pp. 1105--1109.

\bibitem{qiu2025synspeech}
Qifeng Qiu, Yutian Li, Lap-Kei Lee, Fu~Lee Wang, and Zhenguo Yang,
\newblock ``Synspeech: A dataset and benchmark for fake speech detection,''
\newblock in {\em Proceedings of the 7th ACM International Conference on Multimedia in Asia}, 2025, pp. 1--7.

\bibitem{luong2025llamapartialspoof}
Hieu-Thi Luong, Haoyang Li, Lin Zhang, Kong~Aik Lee, and Eng~Siong Chng,
\newblock ``Llamapartialspoof: An llm-driven fake speech dataset simulating disinformation generation,''
\newblock in {\em ICASSP 2025-2025 IEEE International Conference on Acoustics, Speech and Signal Processing (ICASSP)}. IEEE, 2025, pp. 1--5.

\bibitem{Pratap_Xu_Sriram_Synnaeve_Collobert_2020}
Vineel Pratap, Qiantong Xu, Anuroop Sriram, Gabriel Synnaeve, and Ronan Collobert,
\newblock ``Mls: A large-scale multilingual dataset for speech research,''
\newblock {\em Interspeech 2020}, Oct 2020.

\bibitem{Eren_Coqui_TTS_2021}
Gölge Eren and {The Coqui TTS Team},
\newblock ``{Coqui TTS},'' Jan. 2021.

\bibitem{Bain_Huh_Han_Zisserman_2023}
Max Bain, Jaesung Huh, Tengda Han, and Andrew Zisserman,
\newblock ``Whisperx: Time-accurate speech transcription of long-form audio,''
\newblock {\em INTERSPEECH 2023}, Aug 2023.

\bibitem{McAuliffe_Socolof_Mihuc_Wagner_Sonderegger_2017}
Michael McAuliffe, Michaela Socolof, Sarah Mihuc, Michael Wagner, and Morgan Sonderegger,
\newblock ``Montreal forced aligner: Trainable text-speech alignment using kaldi,''
\newblock {\em Interspeech 2017}, Aug 2017.

\bibitem{Ko_Peddinti_Povey_Seltzer_Khudanpur_2017}
Tom Ko, Vijayaditya Peddinti, Daniel Povey, Michael~L. Seltzer, and Sanjeev Khudanpur,
\newblock ``A study on data augmentation of reverberant speech for robust speech recognition,''
\newblock {\em 2017 IEEE International Conference on Acoustics, Speech and Signal Processing (ICASSP)}, p. 5220–5224, Mar 2017.

\bibitem{snyder2015musan}
David Snyder, Guoguo Chen, and Daniel Povey,
\newblock ``Musan: A music, speech, and noise corpus,''
\newblock {\em arXiv preprint arXiv:1510.08484}, 2015.

\bibitem{reddy2021dnsmos}
Chandan~KA Reddy, Vishak Gopal, and Ross Cutler,
\newblock ``Dnsmos: A non-intrusive perceptual objective speech quality metric to evaluate noise suppressors,''
\newblock in {\em ICASSP 2021-2021 IEEE International Conference on Acoustics, Speech and Signal Processing (ICASSP)}. IEEE, 2021, pp. 6493--6497.

\bibitem{Tak_Jung_Patino_Kamble_Todisco_Evans_2021}
Hemlata Tak, Jee-weon Jung, Jose Patino, Madhu Kamble, Massimiliano Todisco, and Nicholas Evans,
\newblock ``End-to-end spectro-temporal graph attention networks for speaker verification anti-spoofing and speech deepfake detection,''
\newblock {\em 2021 Edition of the Automatic Speaker Verification and Spoofing Countermeasures Challenge}, Sep 2021.

\bibitem{ravanelli2018speaker}
Mirco Ravanelli and Yoshua Bengio,
\newblock ``Speaker recognition from raw waveform with sincnet,''
\newblock in {\em 2018 IEEE spoken language technology workshop (SLT)}. IEEE, 2018, pp. 1021--1028.

\bibitem{baevski2020wav2vec}
Alexei Baevski, Yuhao Zhou, Abdelrahman Mohamed, and Michael Auli,
\newblock ``wav2vec 2.0: A framework for self-supervised learning of speech representations,''
\newblock {\em Advances in neural information processing systems}, vol. 33, pp. 12449--12460, 2020.

\bibitem{li2025frame}
Menglu Li, Xiao-Ping Zhang, and Lian Zhao,
\newblock ``Frame-level temporal difference learning for partial deepfake speech detection,''
\newblock {\em IEEE Signal Processing Letters}, 2025.

\bibitem{Liu_Truong_Kumar}
Tianchi Liu, Duc-Tuan Truong, Rohan Kumar~Das, Kong Aik~Lee, and Haizhou Li,
\newblock ``Nes2net: A lightweight nested architecture for foundation model driven speech anti-spoofing,''
\newblock {\em IEEE Transactions on Information Forensics and Security}, vol. 20, pp. 12005–12018, 2025.

\end{thebibliography}
\end{document}